\documentclass[12pt]{article}
\usepackage{graphicx}
\usepackage{amssymb}

\begin{document}

\title{A New Gravitational Lens Candidate with Large Image Separation in the SDSS DR5 Data}

\author{{\small \it Eusebio S\'anchez \'Alvaro,} \\
        {\small \it Francisco Javier Rodr\'iguez Calonge} \\
        ~ \\
        {\small \it CIEMAT} \\
        {\small \it Avda. Complutense 22} \\
        {\small \it E-28040 Madrid (Spain)} \\
       }      
\date{~}
\maketitle
\begin{abstract}
The discovery of a gravitational lens candidate is 
reported. The quasars SDSS J111611.73+411821.5 and
SDSS J111610.68+411814.4 are recognized as two images
of the same object, being strongly lensed by the
closer galaxy SDSS J111611.03+411820.9. The source
is located at a redshift of $z \sim 3$, while the redshift
of the lens galaxy is $z \sim 0.25$. The separation of
the images is large, $\sim 13$ arcsec. Commonly used models
of the mass distribution for the lens galaxy with
values of the parameters in the expected range describe the 
positions and fluxes of the images.  

\end{abstract}
\newpage

\section*{Introduction}

Gravitational lenses are widely used for 
astrophysical and cosmological studies. Individually 
analyzed, they can provide a better understanding of both 
the lens and the source. The lensing effect depends only 
on the distribution of mass within the lens, and not on the 
type of matter, and can be used to study the distribution
of dark matter in galaxy or galaxy cluster 
halos~\cite{ref:darkmatter}. The statistical properties 
of the lens systems are sensitive to the cosmological 
constant~\cite{ref:cosmocons}, the global Hubble 
constant~\cite{ref:hubble}, or the formation and 
evolution of galaxies~\cite{ref:evogal}.

Currently, about 100 lensed quasars have been 
discovered\footnote{http://cfa-www.harvard.edu/castles}, but 
the recent development of 
massive sky surveys is increasing this number quickly. The 
discovery of a new lens candidate in the  
Sloan Digital Sky Survey (SDSS)\footnote{http://www.sdss.org}
 data is reported here. Several gravitational lenses have
been discovered previously in the SDSS 
data~\cite{ref:sdss_lenses}. Collecting
a large sample of gravitational lenses with well-defined
statistical properties is important to obtain 
consistent cosmological and astrophysical conclusions, and 
the SDSS gives a good oportunity to obtain such a sample. 

In the analysis presented here, the  
quasars SDSS J111610.8+411814.4 and SDSS J111611.73+411821.5 
are identified as two images of the same quasar, being lensed 
by the closer galaxy SDSSJ111611.03+411820.9. Both quasars are 
in the spectroscopic catalog of the survey, while the galaxy 
is a photometric object. The system is studied and modelled 
to check the consistency of the gravitational lens hypothesis.

\section*{Observation of the Lens Candidate}

\subsection*{The Sloan Digital Sky Survey}

The SDSS~\cite{ref:sdss} is a large-area imaging survey of the 
north Galactic cap. It has covered nearly a quarter of the Celestial 
Sphere (8000 square degrees). The survey has recorded 
around $1.3\times10^8$ 
objects with well-calibrated photometry~\cite{ref:sdss_photometry}. The 
photometric errors are typically less than 0.03 
magnitude~\cite{ref:sdss_magnitudes}. In addition to that, there 
is a spectroscopic survey, that has collected spectra for roughly
$10^6$ galaxies and $10^5$ quasars, and has covered a total area
of 5713 square degrees using an optimized tiling 
algorithm~\cite{ref:sdss_tiling}.

The survey uses a dedicated telescope~\cite{ref:sdss_telescope} of 
2.5m, located at the Apache Point Observatory, New Mexico. The 
imaging survey 
uses the drift-scan mode of the 142 Mpixel camera~\cite{ref:sdss_camera}
to obtain data in five broad bands~\cite{ref:sdss_filter}
, $u$ $g$ $r$ $i$ $z$, centered at 3561, 4676, 6176, 7494 and 
8873 \AA ~respectively. The images are processed automatically 
by a photometric pipeline~\cite{ref:sdss_pipeline}, and are 
calibrated astrometrically~\cite{ref:sdss_astrometry} and 
photometrically~\cite{ref:sdss_photometry}.

Some objects are selected from the imaging data for spectroscopy
using a variety of algorithms, that include a deeper sample 
of galaxies~\cite{ref:sdss_galaxies} and 
quasars~\cite{ref:sdss_qso}. The spectra are observed by means of 
a pair of double spectrographs able to take 640 spectra 
simultaneously. They are fed by optical fibers each 3'' in 
diameter. The spectra cover the wavelength range from 3800 to 
9200 \AA~, with a resolution of 
$\lambda/\Delta \lambda \sim 2000$.

The results presented here correspond to the Data Release 5 (DR5) 
of the SDSS~\cite{ref:sdss_dr5}. Previous data releases are 
described elsewhere~\cite{ref:sdss_datareleases}. 

\subsection*{Observation of the Lens Candidate}

\begin{figure}[p]
\begin{center}
\includegraphics[width=15.0truecm]{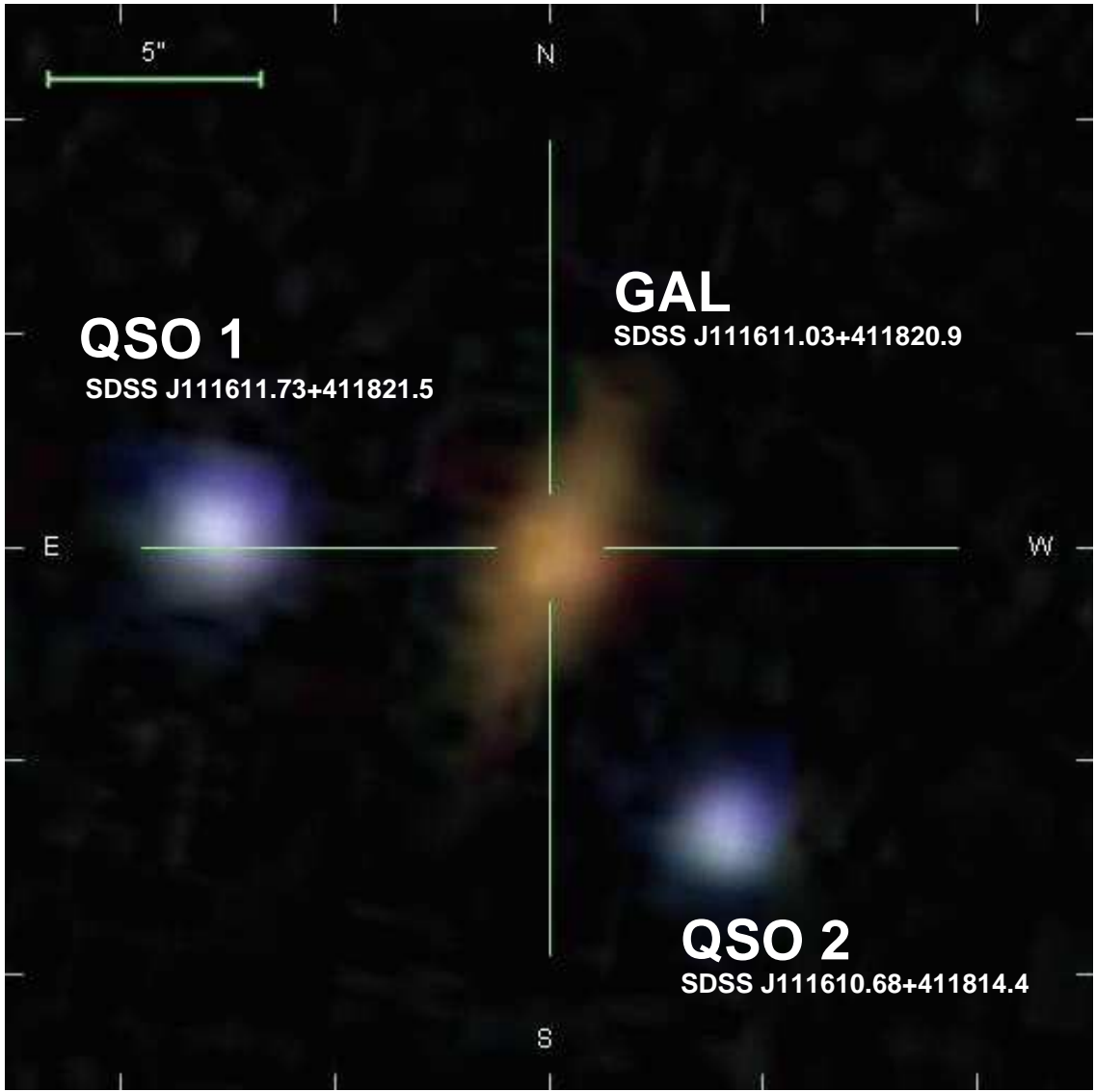}
\caption{SDSS image of the lens candidate. The two images of the
         quasar are compatible with point-like objects, while the
         galaxy is an extended object located in the middle of the
         two quasar images. 
         \label{fig:lensimage}}
\end{center}
\end{figure}

The identification of the lens candidate presented here
was performed during a study of close quasar pairs. The SDSS image 
of the system can be seen in Figure~\ref{fig:lensimage}, where
the two images of the quasar are labelled QSO 1 and QSO 2, and are 
compatible with point-like objects, and the lens galaxy is clearly 
visible between them as an extended object, labelled GAL. The large 
separation between the images makes the system a very clear 
candidate, since the objects are perfectly resolved even in the SDSS 
spectroscopic sample, and no additional follow-up is necessary. The
main astrometric and photometric features of the objects are 
described in Table~\ref{tab:features}.

\begin{figure}[p]
\begin{center}
\includegraphics[width=16.0truecm]{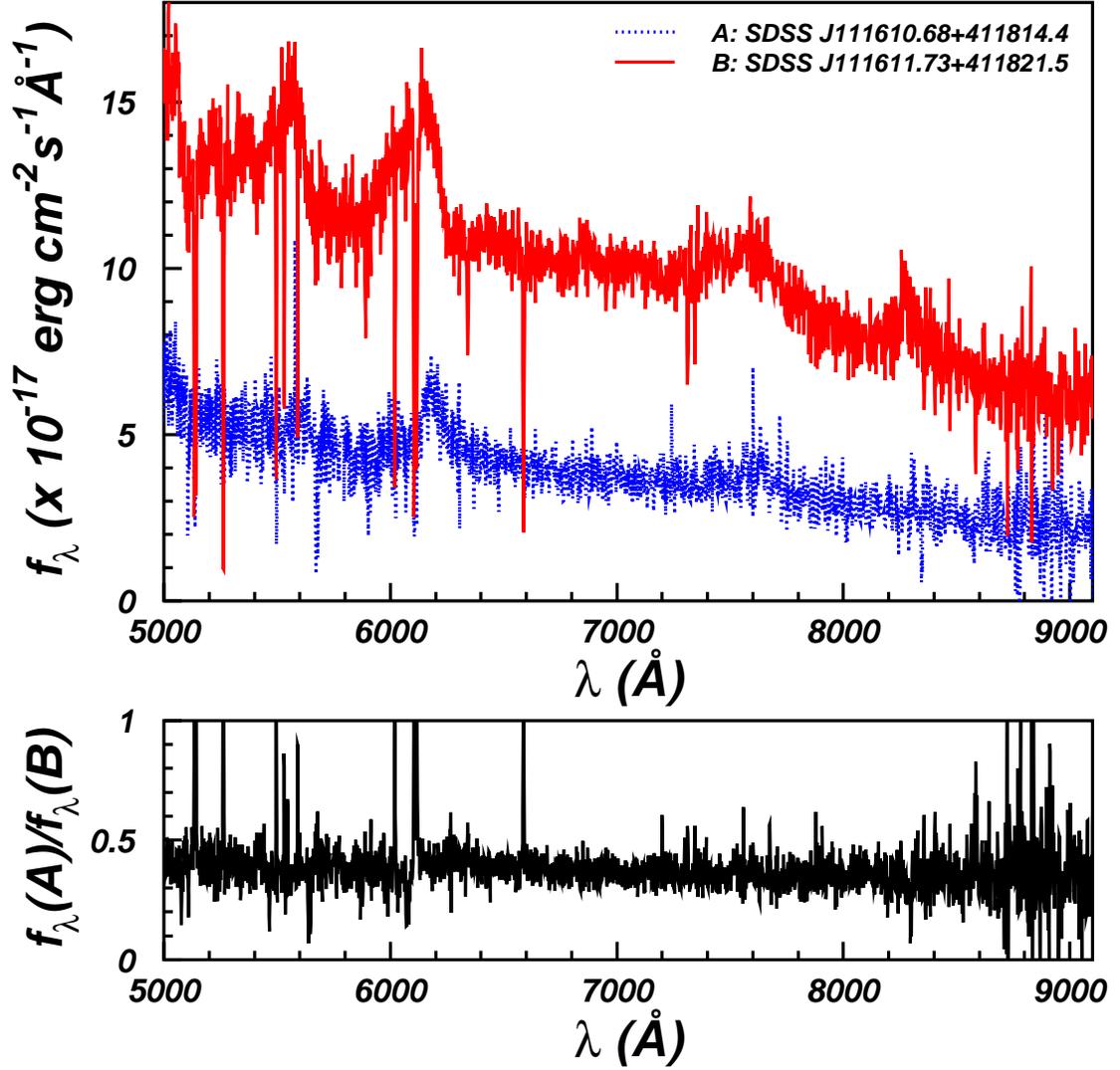}
\caption{Top: Spectra of the two QSO images taken in the
         SDSS spectroscopic survey. Bottom: Ratio of both 
         spectra. The value is constant along the considered
         wavelenghts range.
         \label{fig:qsospectra}}
\end{center}
\end{figure}

The two images of the quasar are included in the spectroscopic 
sample of the SDSS. Their spectra can be seen in 
Figure~\ref{fig:qsospectra}-top. The angular separation between
the two quasar images is large, $13.7''$, and the separation
of each image and the lens galaxy is $7.6''$ and $7.9''$. The observed
redshifts of the two images using the CIV line are $3.01\pm0.04$ and
$3.01 \pm 0.03$. The optimal redshifts from the SDSS spectroscopic
calculations are $2.971\pm0.003$ and $2.980 \pm 0.005$. The redshifts
are compatible within errors. 

Moreover, the shapes of both spectra are very similar. The ratio of 
spectra is shown in Figure~\ref{fig:qsospectra}-bottom, and it is 
constant along the full range of considered wavelengths. Considering 
both quasars as lensed images of a single object is then supported both 
by the similarity of the spectra and by the consistency of the redshifts.

There is no spectrum of the lens galaxy, which is located in the
middle of the two images of the
quasar, but the photometric redshift computed 
for it in the SDSS data is $z=0.18 \pm 0.03$ using the templates
algorithm~\cite{ref:photoz} or $z=0.25 \pm 0.03$ using the
modified neural network method~\cite{ref:photoz2}. The two values
are only marginally compatible, but the lens system is very
weakly dependent on this redshift. The most sensitive quantity
is the time delay between images, that has not been explored in
this analysis. For the other parameters, the full analysis has 
been performed using both values of the redshift and the conclusions 
are the same, with compatible values for all the parameters. The 
numbers quoted in the text are for a redshift of 0.25.

\section*{Lens Model}

The gravitational lens hypothesis is analysed using 
two mass models, a Singular Isothermal Ellipsoid (SIE) and a 
Singular Isothermal Sphere with shear (SISS). The {\sf gravlens/lensmodel} 
software~\cite{ref:lensmodel} is used. The 
assumed errors are $\pm 0.1''$ in the positions and 20\% 
in the fluxes. These errors are larger than the actual errors
on the measurements, and are adopted to account for possible
external perturbations. Both models have 8 parameters (the 
lens position, the Einstein
radius, the ellipticity or shear, the position angle, the source position
and the source flux) and the system offers 8 constraints 
(the image positions, the lens position and the fluxes of the images), what 
means that the data should be described perfectly in any fit to these 
models, since there are no degrees of freedom. However, sensible 
values for the parameters should be obtained from the fits.

\begin{figure}[p]
\begin{center}
\begin{tabular}{c}
\includegraphics[width=8.0truecm]{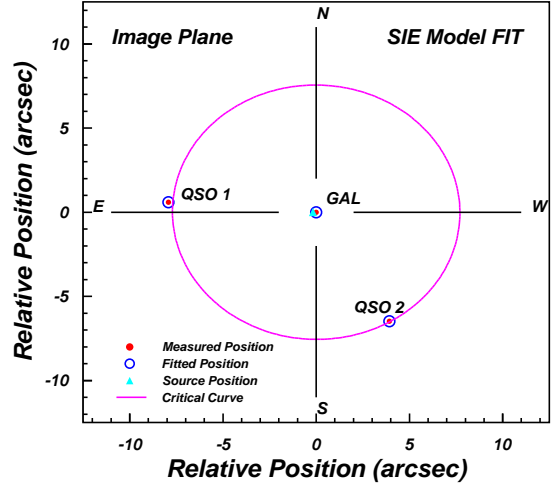} \\
\includegraphics[width=8.0truecm]{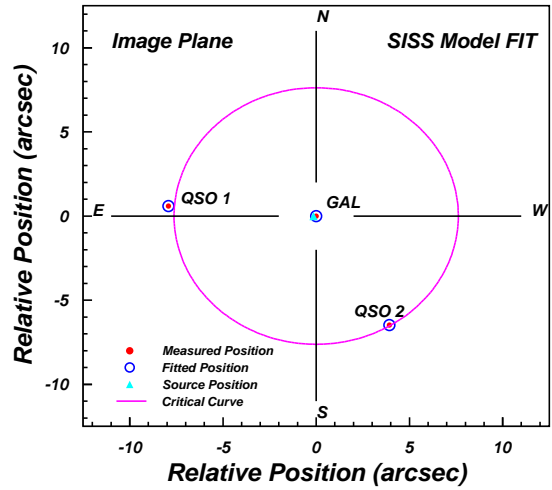}\\
\end{tabular}
\caption{Measured (dots) and fitted (open circles) positions 
         of the QSO images and the lens galaxy for the two
         mass models considered, SIE (top) and SISS (bottom). The 
         size of the dots and the open circles does not correspond 
         to the errors, and is chosen to make them visible. The critical
         curves for the models and the fitted positions of the
         source are also plotted.
         \label{fig:lensangles} }
\end{center}
\end{figure}

The measured positions of the images relative to the central galaxy
in the lens plane are shown in Figure~\ref{fig:lensangles} as dots, while 
the fitted positions are shown as open circles. Both models, SIE and SISS
are able to describe the observed positions, as expected. Also the fitted
source position and the critical curves for the lens system are presented.

The parameters for the lens obtained from each model are shown in 
Table~\ref{tab:siesiss}.

The mass of the lens galaxy, defined as all the mass contained inside
the Einstein radius, is $M_L = (5.3 \pm 0.1) \times 10^{12} M_{\odot}$ in
both models.

The fitted ellipticity for the lens galaxy in the SIE model
is smaller than the 
measured ellipticity for the light, which is around 0.3 in all
the bands, as presented in Table~\ref{tab:lensgalaxy}. The fitted 
shear in the SISS model is in the typical range for lens systems. The 
fitted position angles are not compatible with the measured angle for the
light, shown in Table~\ref{tab:lensgalaxy}, what means that some slight 
external perturbation may be present. In fact, there
seems to be a group or cluster of galaxies between the source and the
lens galaxy, at $z \sim 0.5$, that may also contribute to the lensing.

Since the redshifts for the source quasar and the lens galaxy
are known, it is possible to compute the angular diameter 
distances, for a flat cosmology with 
$\Omega_{\Lambda}=0.7$, $\Omega_{M}=0.3$ and $h=0.71$. Then, the 
expected magnifications of the quasar images can be obtained from 
the lens models. The expected ratio of fluxes of the two images of the
quasar is found to be $0.38 \pm 0.02$ for both models, in agreement with 
the observed value, $0.38 \pm 0.01$, as expected. The observed value
has been obtained from a fit to the ratio of spectra
of Figure~\ref{fig:qsospectra}-bottom and confirmed from the differences 
of magnitudes from Table~\ref{tab:features}.

The velocity dispersion of the mass producing the lens in the SIE
model can be determined from the Einstein radius, and it is found
to be $\sigma = 560$ km/s, higher than expected for a typical
elliptical galaxy. This fact is not strange, since the mass
distribution producing the lens extends for $5.80 \pm 0.11$ effective radii
from the center of the galaxy. The effective radius of the galaxy, taken
as the radius that contains half of the petrosian flux of the
galaxy in the r-band is read from the SDSS database. Then, using 
differents mass distributions (described in {\sf gravlens/lensmodel}), the
mass inside the effective radius can be inferred from the mass
contained inside the Einstein radius. When the  SIE model is used to 
evaluate the mass of the galaxy inside the effective radius, the 
obtained value is 
$M_{R_{eff}}^{SIE}=(0.91 \pm 0.02) \times 10^{12} M_{\odot}$. The
corresponding velocity dispersion for this radius is 
$\sigma_{R_{eff}}=232$ km/s. These are typical values for an elliptical 
galaxy. If a NFW profile is used, the mass contained in the
effective radius is 
$M_{R_{eff}}^{NFW}=(2.33 \pm 0.05) \times 10^{12} M_{\odot}$. 

An extended part of the dark
matter halo is contributing to the lens effect. The 
contribution of the halo is confirmed by the fact that the 
mass-to-luminosity ratio of the lens galaxy is 
$M_{L}/L_{L} \sim 170 M_{\odot}/L_{\odot}$. This
value is in the expected range if the dark matter halo 
contributes to the lens. The absolute luminosity of the
galaxy has been obtained from the SDSS database. The 
mass-to-luminosity ratios inside the effective
radius, are $\sim 30 M_{\odot}/L_{\odot}$ for the isothermal
models and $\sim 100 M_{\odot}/L_{\odot}$ for the NFW \
model. The value for
the NFW is much higher than expected for a typical elliptical
galaxy. The value for the isothermal profile is high, but 
can be still in the expected range for elliptical galaxies.

\section*{Conclusions}

A candidate to strong lensed quasar by a closer galaxy has
been found in the SDSS DR5 data. The system fulfills the
conditions to be considered a gravitational lens. There
are two point-like images with very similar redshifts ($\sim 3$)
and spectra and there is a galaxy between the images with a 
redshift smaller than the images ($\sim 0.25$). The angular 
separation between the two images is large($\sim 13 ''$)

The fitted Einstein radius for the lens is 
large, $\theta_{E} = 7.68 \pm 0.09 ''$. The total mass inside 
the Einstein radius is $M_{L}=(5.3 \pm 0.1) \times 10^{12} M_{\odot}$. The 
mass-to-luminosity ratio of the lens galaxy at the Einstein radius, or
5.8 effective radii, is high, $M_L/L_{L}>100 M_{\odot}/L_{\odot}$. This 
is compatible with a substantial contribution of the dark matter halo
to the lensing mass. The extrapolation of the mass value from the 
Einstein radius to the effective radius with different mass 
distributions shows that the isothermal models are a good 
description for an elliptical galaxy, with mass and velocity
dispersion in the expected range. The orientation angle resulting from the
fit (in both models) is misaligned with respect to the 
orientation angle of light, which may indicate some small 
contribution of an external perturbation. In fact, there seems to be
a group or cluster of galaxies at $z \sim 0.5$ that may have some
contribution to the lensing of the quasar.

\section*{Acknowledgments}

Funding for the Sloan Digital Sky Survey (SDSS) and SDSS-II 
has been provided by the Alfred P. Sloan Foundation, the 
Participating Institutions, the National Science Foundation, the 
U.S. Department of Energy, the National Aeronautics and Space 
Administration, the Japanese Monbukagakusho, and the Max Planck 
Society, and the Higher Education Funding Council for England. The 
SDSS Web site is http://www.sdss.org/.

The SDSS is managed by the Astrophysical Research Consortium (ARC) 
for the Participating Institutions. The Participating Institutions 
are the American Museum of Natural History, Astrophysical Institute 
Potsdam, University of Basel, University of Cambridge, Case Western 
Reserve University, The University of Chicago, Drexel 
University, Fermilab, the Institute for Advanced Study, the Japan 
Participation Group, The Johns Hopkins University, the Joint Institute 
for Nuclear Astrophysics, the Kavli Institute for Particle Astrophysics 
and Cosmology, the Korean Scientist Group, the Chinese Academy of 
Sciences (LAMOST), Los Alamos National Laboratory, the 
Max-Planck-Institute for Astronomy (MPIA), the Max-Planck-Institute 
for Astrophysics (MPA), New Mexico State University, Ohio State 
University, University of Pittsburgh, University of Portsmouth, Princeton 
University, the United States Naval Observatory, and the University 
of Washington.

\newpage
\begin{table}[htb]
\begin{center}
\begin{tabular}{cccc}
\hline \hline
Object            & GAL              & QSO 1            &  QSO 2           \\
\hline
$\Delta$ra ($''$) & 0.0              & 10.541           &  0.590           \\
$\Delta$dec ($''$)& 0.0              & -5.227           & -6.473           \\
\hline
$u$               & $21.75 \pm 0.28$ & $20.40 \pm 0.05$ & $21.36 \pm 0.11$ \\
$g$               & $19.79 \pm 0.02$ & $18.54 \pm 0.01$ & $19.45 \pm 0.01$ \\
$r$               & $18.62 \pm 0.01$ & $18.15 \pm 0.01$ & $19.15 \pm 0.01$ \\
$i$               & $18.03 \pm 0.01$ & $17.93 \pm 0.01$ & $19.00 \pm 0.01$ \\
$z$               & $17.64 \pm 0.03$ & $17.96 \pm 0.02$ & $19.03 \pm 0.04$ \\
\hline \hline
\end{tabular}
\end{center}
\caption{\label{tab:features} Astrometry and photometry of the lens objects. The
difference in right ascension and declination with respect to the lens galaxy is
given in arcsec. The coordinates of the galaxy are 
ra$=169.045962^{\circ}$, dec$=41.305817^{\circ}$. The given magnitudes are
the SDSS model magnitudes in the five bands.}
\end{table}

\begin{table}[htb]
\begin{center}
\begin{tabular}{cccccc}
\hline \hline
Model &  $\theta_{Einstein}$ ($''$) & Mass ($10^{12} M_{\odot}$) & Shear & Ellipticity & Angle ($^{\circ}$) \\
\hline
SIE  & $7.68 \pm 0.07$ & $5.3 \pm 0.1$ &  --  & $0.06 \pm 0.02$ & $263 \pm 6$ \\
SISS & $7.68 \pm 0.07$ & $5.3 \pm 0.1$ & $0.02 \pm 0.01$ &  --  & $263 \pm 6$ \\
\hline \hline
\end{tabular}
\end{center}
\caption{\label{tab:siesiss} Lens parameters obtained from 
fits to two models of the mass distribution: SIE and 
SISS. Ellipticity or shear plus orientation angle may 
indicate some influence of external perturbations.}
\end{table}

\begin{table}[htb]
\begin{center}
\begin{tabular}{cccc}
\hline \hline
Band & Ellipticity & Orientation Angle ($^{\circ}$) \\
\hline
$u$  & $0.41 \pm 0.73$ & 142 \\
$g$  & $0.34 \pm 0.06$ & 156 \\
$r$  & $0.28 \pm 0.03$ & 157 \\
$i$  & $0.30 \pm 0.03$ & 153 \\
$z$  & $0.26 \pm 0.06$ & 157 \\
\hline \hline
\end{tabular}
\end{center}
\caption{\label{tab:lensgalaxy} Main physical characteristics
of the lens galaxy in the five SDSS bands. The ellipticity and
the orientation angle are the De Vacouleurs values read from 
the SDSS database. The orientation
angle is defined in the SDSS standard way, east of north.}
\end{table}

\newpage

\newpage

\newpage

\end{document}